\pdfoutput=1
\documentclass[11pt]{article}

\usepackage[preprint]{acl}

\usepackage{times}
\usepackage{latexsym}

\usepackage[T1]{fontenc}
\usepackage[utf8]{inputenc}

\usepackage{microtype}

\usepackage{inconsolata}

\usepackage{graphicx}

\usepackage{enumitem}
\usepackage{multirow}
\usepackage{subcaption}
\title{Lucia: A Temporal Computing Platform for Contextual Intelligence}

\author{
 \textbf{Weizhe Lin\textsuperscript{1}},
 \textbf{Junxiao Shen\textsuperscript{1}}
\\
  \{weizhe.lin, shawn.shen\}@openinterx.com
 \\
 \textsuperscript{1}OpenInterX Research, United States
}

\begin{document}
\maketitle
\begin{abstract}
The rapid evolution of artificial intelligence, especially through multi-modal large language models, has redefined user interactions, enabling responses that are contextually rich and human-like. As AI becomes an integral part of daily life, a new frontier has emerged: developing systems that not only understand spatial and sensory data but also interpret temporal contexts to build long-term, personalized memories. This report introduces Lucia, an open-source Temporal Computing Platform designed to enhance human cognition by capturing and utilizing continuous contextual memory. Lucia introduces a lightweight, wearable device that excels in both comfort and real-time data accessibility, distinguishing itself from existing devices that typically prioritize either wearability or perceptual capabilities alone. By recording and interpreting daily activities over time, Lucia enables users to access a robust temporal memory, enhancing cognitive processes such as decision-making and memory recall.
\end{abstract}

\section{Introduction}
The rapid advancements in artificial intelligence, particularly the emergence of multi-modal large language models (LLMs)~\cite{achiam2023gpt, wang2024qwen2, anthropic2024claude, team2023gemini} like GPT-4~\cite{achiam2023gpt}, have dramatically transformed our interaction with technology. These models exhibit an unprecedented ability to understand and generate human-like language, process visual and auditory information, and interpret 3D spatial environments~\cite{zhao2023survey, yin2023survey, engel2023project}. However, as we push the boundaries of AI, a new frontier emerges: Temporal Computing—the understanding and utilization of time to construct contextual memory that enhances human cognition. This evolution has paved the way for devices that are not only intelligent but also temporally aware, deeply personalized, and seamlessly integrated into our daily lives.

As technology becomes an extension of ourselves, there is a growing demand for AI devices that comprehend and adapt to individual user contexts over time. By leveraging contextual memory, these personalized, temporally-aware, and human-oriented AI devices promise to elevate our interaction with the digital world. They provide intuitive, natural, and efficient user experiences by anticipating needs, understanding preferences, and offering timely assistance without being intrusive, fundamentally enhancing the way we live and work.
\begin{figure*}[!htp]
    \centering
    \includegraphics[width=0.8\textwidth]{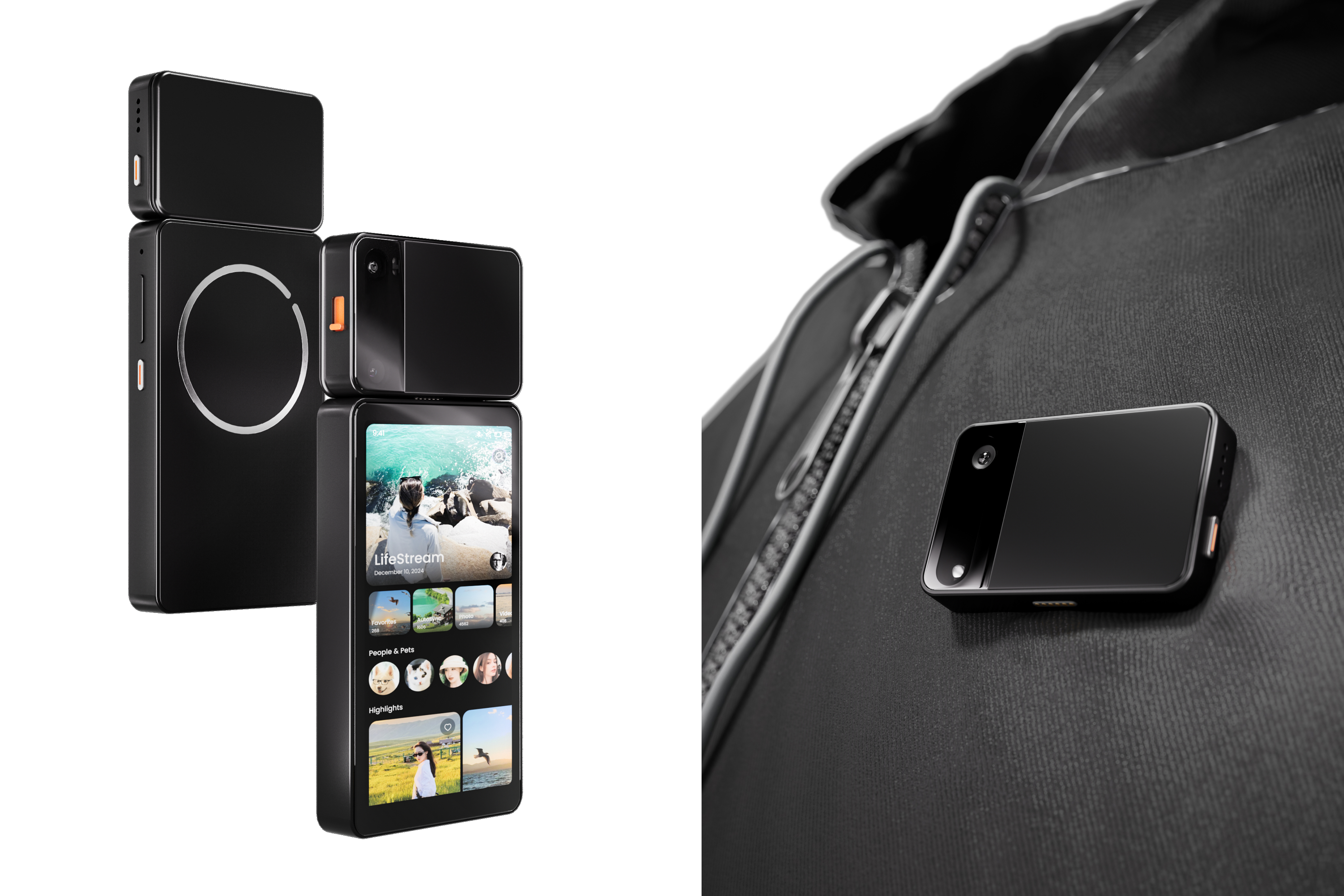}
    \caption{The illustration of the device. The device comprises two main components: a temporal pin designed for multi-modal sensing and a temporal hub that performs computational processing.}
    \label{fig:camera_illustration}
\end{figure*}

In response to this demand, we introduce Lucia, an open-source, publicly available Temporal Computing Platform designed to enhance human cognition through contextual memory. This novel portable device effortlessly attaches to users, efficiently recording their daily activities over time. By leveraging advanced AI capabilities, the device captures long-term activities, creating a rich temporal contextual memory that users can query at any time. Acting as an intelligent assistant or companion, it not only answers queries regarding stored experiences but also understands temporal patterns, thereby augmenting memory recall, enhancing decision-making, and boosting personal productivity.

Lucia builds upon the concepts introduced by Project Aria~\cite{engel2023project}, Meta's all-day wearable AR glasses developed as data collection tools for spatial computing. While Project Aria aims to shift computing paradigms by blending digital interactions into the 3D world through spatial computing, Lucia extends these ideas by emphasizing the temporal dimension. It prioritizes the continuous capture and intelligent interpretation of user activities over time while enhancing practical usability:
Lucia creates a device that not only records but also understands and provides insightful responses based on the user's temporal experiences.
By harnessing the power of modern AI models and Temporal Computing principles, it creates a device that not only records but also understands and provides insightful responses based on the user's temporal experiences. Moreover, unlike many devices that require frequent battery changes or tethering to external power packs for continuous operation, Lucia serves users all day without such inconveniences, offering a seamless and unobtrusive user experience.

\begin{figure}[!h]
    \centering
    \includegraphics[width=\linewidth]{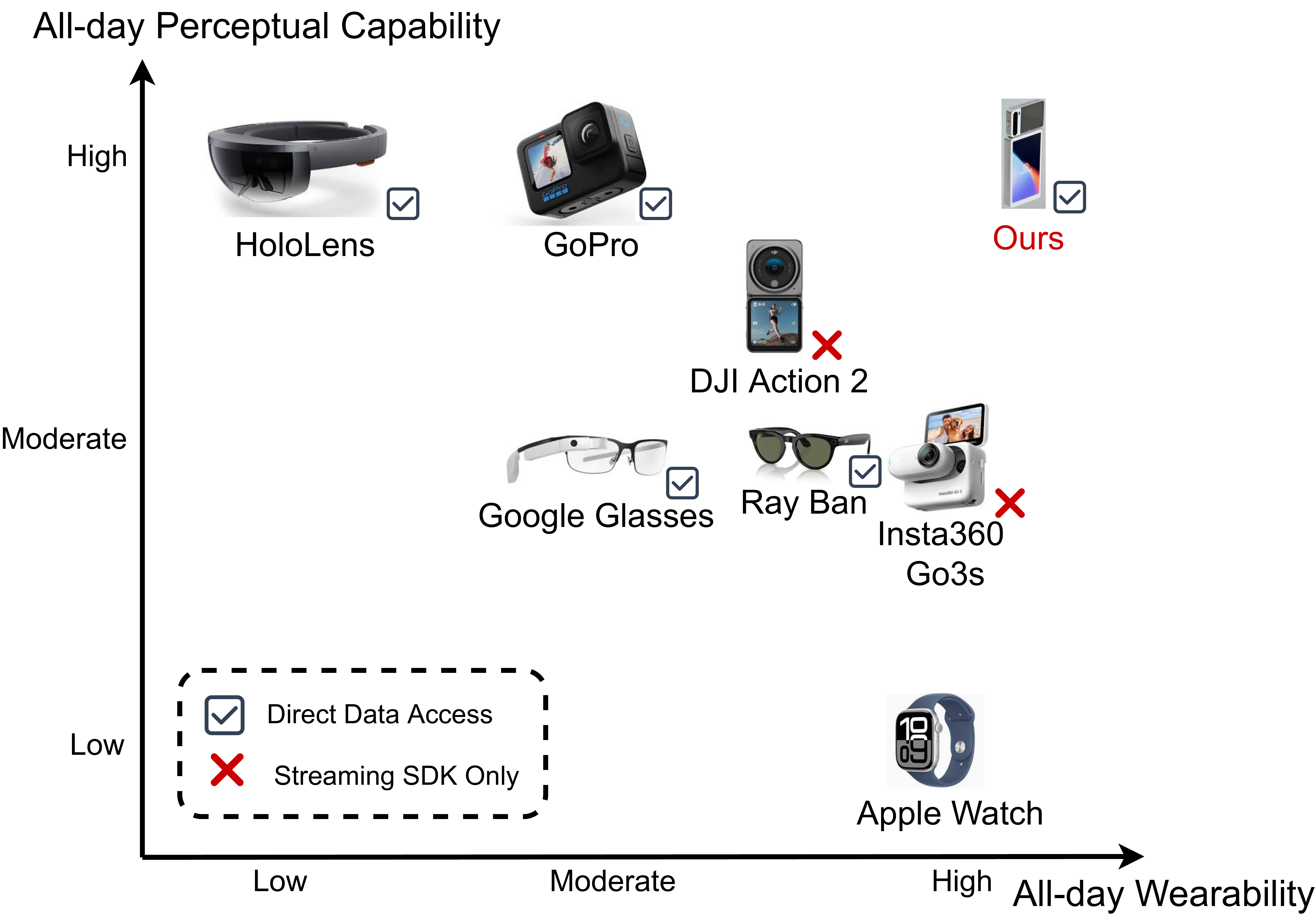}
    \caption{Comparison of Devices Based on All-day Wearability and Perceptual Capability. The X-axis represents All-day Wearability, indicating how comfortable and unobtrusive a device is for continuous, all-day use without the user sensing its presence. The Y-axis represents All-day Perceptual Capability, measuring the device's ability to continuously record and process data throughout the day, including direct real-time access to recorded data. Lucia is positioned in the upper-right quadrant, showcasing its superior performance in both dimensions. Other devices are plotted according to their relative capabilities, highlighting the trade-offs between wearability and perceptual functions in existing technologies such as Apple Watch, GoPro, Insta360 GO3s, DJI Action 2, Meta Ray-Ban, Google Glasses, and Microsoft HoloLens.}
    \label{fig:hardware_comparsion}
\end{figure}

To further illustrate the unique advantages of Lucia as an all-day wearable perceptual computer, we compare it with existing devices in the market in Fig.~\ref{fig:hardware_comparsion}.
In this comparison, we focus on two key hardware aspects: all-day wearability and all-day perceptual capability. All-day wearability refers to the comfort and unobtrusiveness of a device for continuous, all-day use without the user sensing its presence and without the need for frequent charging. All-day perceptual capability measures a device's ability to continuously record and process data throughout the day, assuming that wired charging or battery replacement is available. 
Note that this comparison focuses solely on the hardware capabilities of these devices, excluding the assessment of their AI functionalities.

Many current devices either excel in wearability or perceptual capabilities but rarely both, and often lack the capacity for real-time data access essential for Temporal Computing. 
% For instance, the Apple Watch offers excellent wearability but lacks comprehensive perceptual functions like continuous video recording. Devices like GoPro provide high perceptual capabilities but are not designed for comfortable, unobtrusive all-day wear due to their relatively high weight. The Insta360 GO 3 boasts high wearability but only moderate perceptual capability due to limitations like battery life, inability to operate all day even with wired charging or battery replacement, and the restriction of only providing a streaming SDK, which prohibits direct real-time access to recorded data. Similarly, the DJI Action 2 offers moderate to high wearability and perceptual capabilities, with the advantage of interchangeable batteries for extended use. However, it also only provides a streaming SDK, limiting direct real-time data access.
Lucia uniquely combines high all-day wearability and high all-day perceptual capability, along with direct real-time access to recorded data, positioning it at the forefront of Temporal Computing devices. 
% Figure 1 illustrates this comparison, placing Lucia in the upper-right quadrant to showcase its superior performance in both dimensions.

This technical report demonstrates the necessity and efficacy of our new device, exploring its design principles, functionality, and the significant impact it holds for the future of human-computer interaction. By enhancing human cognition through contextual memory, we offer a step towards a more interconnected and intelligent way of living. We have made these resources available to research institutions around the world to foster advancements in the field of personalized AI. We aim to inspire further innovation and collaboration in this rapidly evolving domain.

\section{Lucia}

\subsection{Project Objectives and Goals} 
Lucia is designed to transform personal data interaction and environmental awareness by creating an open-source, accessible temporal computing platform dedicated to contextual intelligence applications. Its primary goal is to develop a wearable device that seamlessly integrates into users' daily lives, providing a platform to test and validate various contextual intelligence functionalities.

To realize this vision, the device must fulfill three essential criteria. First, it should attach effortlessly to users, ensuring minimal intrusion and maximum convenience. Second, it must support continuous video recording to capture daily activities efficiently, managing storage resources without compromising data integrity. These features enable all-day data capture, essential for developing applications that rely on unobtrusive, continuous data collection throughout daily life. Third, it should allow users to interact with their stored experiences through natural language queries, enhancing cognitive processes such as memory recall and decision-making. This feature supports seamless, real-time interactions with the device’s contextual intelligence applications, delivering timely and relevant responses.

By meeting these objectives, Lucia aims to bridge the gap between personal data collection and meaningful, user-centered applications, setting the stage for smarter and more interconnected living. The project makes this platform available to research institutions worldwide, fostering innovation in personalized AI technologies and promoting an open-source community dedicated to advancing contextual intelligence applications.

\begin{figure}[h]
    \centering
    \includegraphics[width=0.3\linewidth]{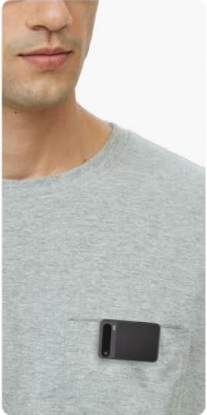}
    \caption{The temporal pin of Lucia is designed for a nearly imperceptible user experience, weighing only 44 grams compared to the 220 grams or more typical of standard smartphones.}
    \label{fig:camera_photo}
\end{figure}

\subsection{Open-Source Platform and Public Availability}

Lucia is founded on openness and collaboration, leveraging an open-source model to invite developers, researchers, and enthusiasts to shape its growth. The source code, documentation, and development tools are publicly accessible in a repository for anyone interested. The project also fosters partnerships with universities and research organizations, enhancing its technological scope and exploring new applications. To broaden accessibility, the physical device will be available through direct purchase and partnerships, while regular software updates provide users with the latest features. By operating transparently, the project builds trust through clarity in data handling and algorithm functionality.

\section{Device} \label{sec:device}

\subsection{Sensor Suite}

\begin{figure*}[!t] 
    \centering
    \begin{subfigure}[b]{0.2\textwidth}
        \centering
        \includegraphics[width=\textwidth]{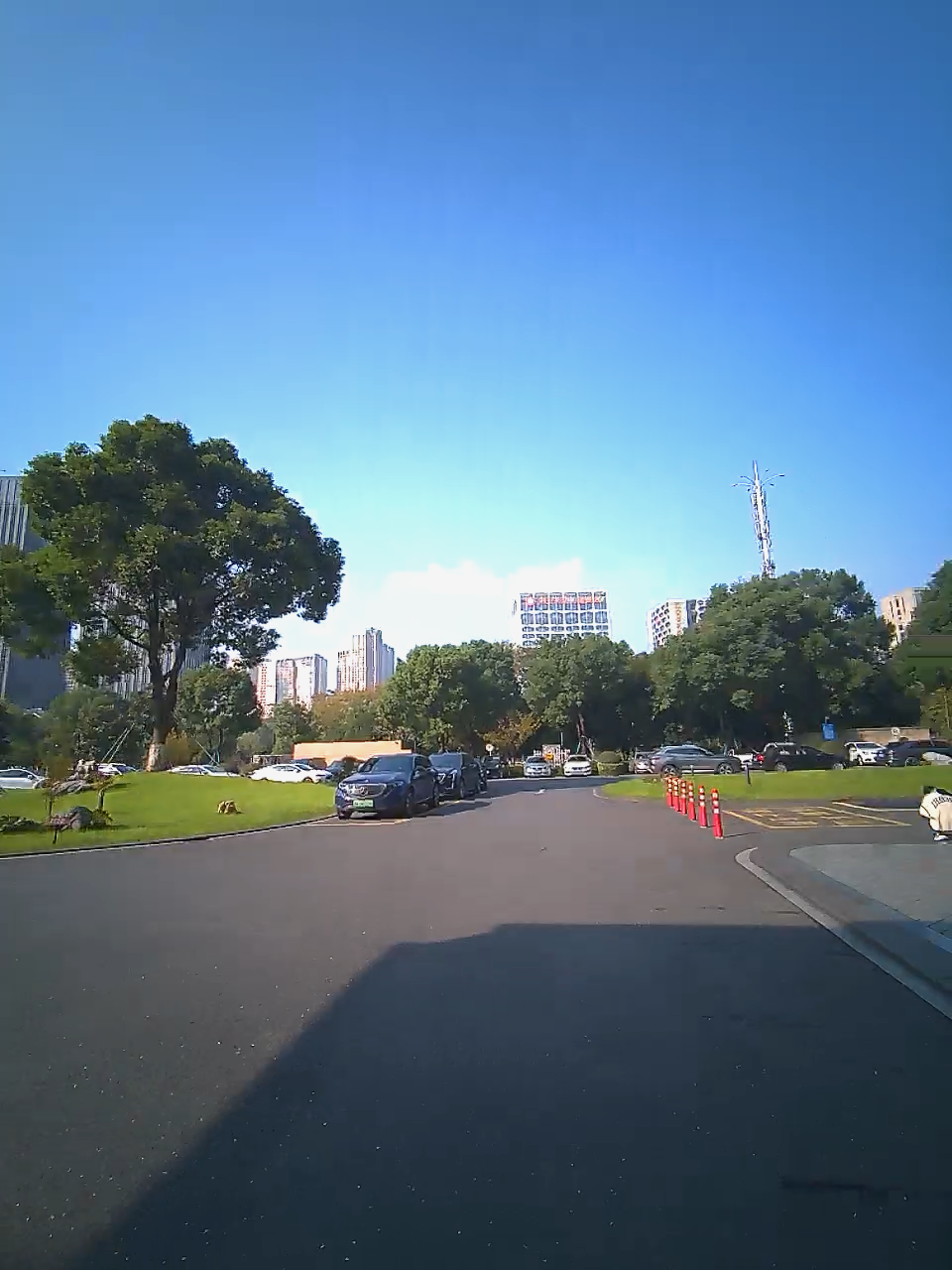}
    \caption{Sample images from the camera. It can record users' daily experiences at high resolution efficiently.} \label{fig:camera_samples} 
    \end{subfigure}%
    \hfill
    \begin{subfigure}[b]{0.78\textwidth}
        \centering
        \includegraphics[width=\textwidth]{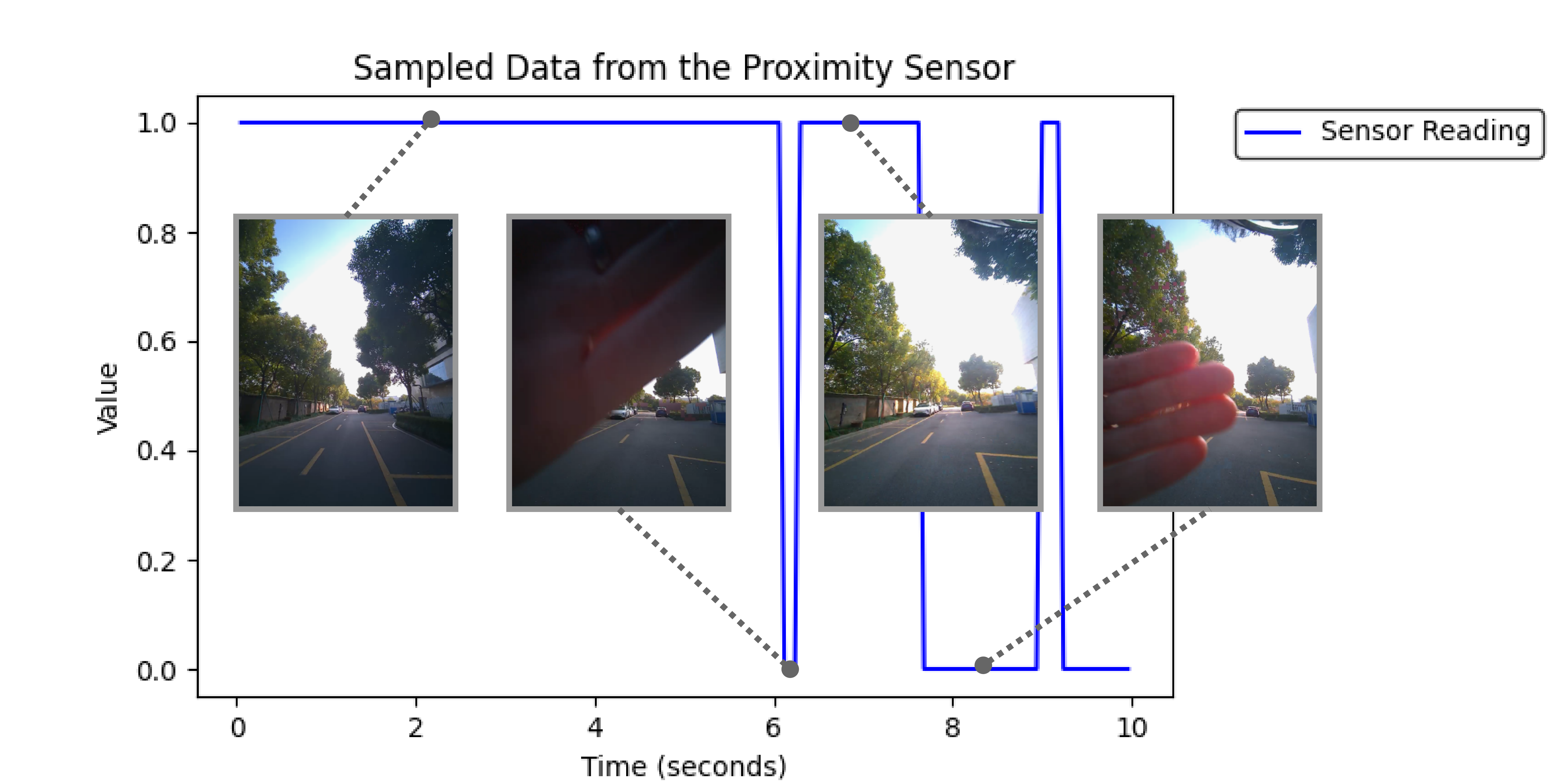}
        \caption{Sample data from the proximity sensor. Combining the sensor and the camera enables more applications such as hand gesture recognition.}
        \label{fig:proximity_samples}
    \end{subfigure}
    \caption{Overview of sample data from the camera and the proximity sensor.}
    \label{fig:combined_samples_1}
\end{figure*}
% \begin{figure*}[!t] 
%     \centering
%     \includegraphics[width=0.5\textwidth]{example-image-c}
%     \caption{Sample data from the IMU sensor. } \label{fig:IMU_samples} 
% \end{figure*}

% \begin{figure*}[!t] 
%     \centering
%     \includegraphics[width=0.5\textwidth]{example-image-a}
%     \caption{Sample data from the dual-microphone array. } \label{fig:audio_samples} 
% \end{figure*}

\begin{figure}[!t] 
    \centering
    \includegraphics[width=1.1\linewidth]{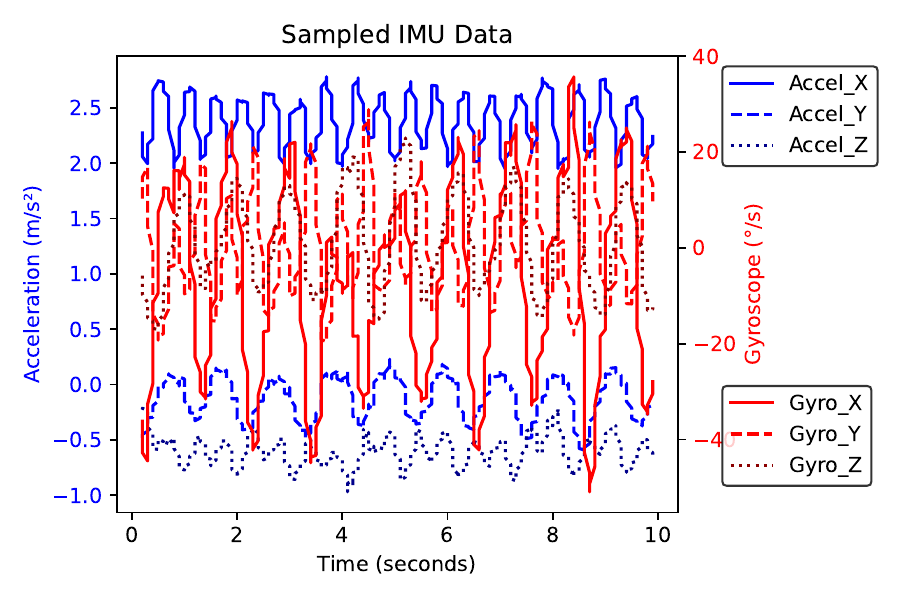}
    \caption{Sample data from the IMU sensor.}
    \label{fig:IMU_samples}
\end{figure}

\begin{figure}[!t]
    \centering
    \includegraphics[width=\linewidth]{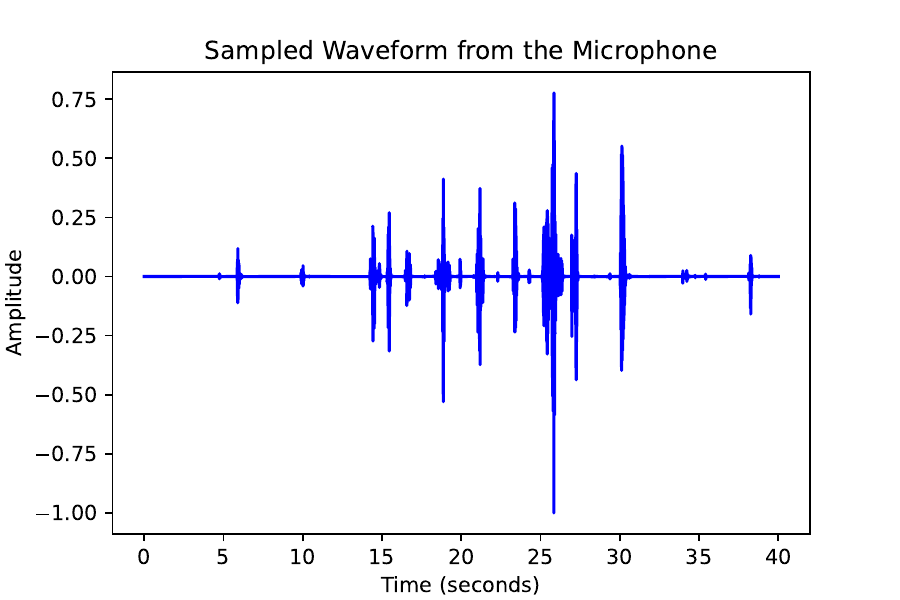}
    \caption{Sample data from the dual-microphone array.}
    \label{fig:audio_samples}
\end{figure}

Lucia is engineered as an advanced wearable platform for capturing multi-modal data, integrating a comprehensive range of sensors tailored for machine perception applications. By emulating the requirements of future wearable hardware, Lucia prioritizes calibrated sensor data collection over real-time processing, thereby facilitating downstream machine learning and perception tasks.

The sensor suite in Lucia offers extensive coverage across visual, motion, and environmental modalities. Each sensor stream is meticulously calibrated and time-synchronized to enhance reliability and accuracy for machine learning applications. The following are the key components of the sensor suite:

\begin{itemize} \setlength\itemsep{0em} \item \textbf{Camera}: Equipped with a 12-megapixel sensor (IMX681) and an adjustable field of view (FOV) of 94°x77°, the camera supports resolutions up to 3840x2160 pixels for photo capture and a range of video formats, including 2016x1512 and 1920x1080 pixels at frame rates up to 30 fps. This allows for high-quality visual data acquisition suitable for various computer vision tasks.

\item \textbf{Inertial Measurement Unit (IMU)}: The device integrates the BMI270 IMU, a 6-axis smart, low-power unit supporting high sampling rates. This IMU measures both accelerative and rotational movements, contributing critical motion data for applications such as Visual SLAM.

\item \textbf{Microphones}: A dual-microphone array enables spatial audio capture, designed to record high-fidelity auditory information across diverse environmental conditions. This setup allows for noise reduction and directional audio processing, enhancing applications in speech recognition and sound localization.

\item \textbf{Proximity Sensor:} The proximity sensor covers a 0–200 mm range in front of the camera, enabling a variety of device control options. This range facilitates gesture recognition, allowing users to interact with the device quickly without needing to access the hub or mobile phone.

\item \textbf{Additional Sensors}: Alongside visual and motion sensors, an integrated ambient light sensor dynamically adjusts camera exposure settings and screen brightness based on surrounding light conditions. 
This optimizes image quality and enhances user experience in varying lighting environments.

\end{itemize}

\subsection{Form Factor and Fit}

As shown in Fig.~\ref{fig:camera_illustration}, the device comprises two primary components: a portable camera module and a hub, connected via a magnetic attachment. The majority of sensors are mounted on the camera module, which is designed to be worn by the user. The camera module combines a robust set of sensors within a compact and lightweight form factor, tailored for extended wearability.

The device employs a magnetic attachment, allowing users to easily detach and assemble the camera module and hub. This attachment supports both wireless charging and data transmission. Magnets are positioned at the back and bottom of the device, facilitating potential future extensions.

The design accommodates various user profiles through adjustable fittings, making it suitable for a diverse range of applications and user demographics. This form factor achieves an optimal balance between weight, size, and sensor integration, providing an adaptable and efficient wearable platform. Weighing only 44 grams, the camera module is unobtrusive, allowing for all-day wear without causing user fatigue or discomfort.

\subsection{Battery Life and Power Management} 
The device is engineered with a high-capacity, rechargeable battery system designed to support continuous multi-modal data recording. Under standard operating conditions, the camera module offers up to 160 minutes of operation, while the hub extends this to up to 400 minutes. Power management in Lucia is finely tuned to balance performance with energy efficiency, enabling users to adjust sensor activation and recording profiles to optimize battery life according to their needs.

Power optimization features include adaptive sensor sampling rates and selective sensor activation, which can be customized based on task requirements. For instance, users can set lower sampling rates for non-critical sensors or disable sensors when not needed. 
This modular power management allows Lucia to accommodate extended sessions, making it well-suited for field data collection where continuous power may be unavailable.

To further enhance efficiency, Lucia incorporates on-demand power-down and sleep modes that can be triggered during periods of inactivity, significantly reducing battery drain. These features ensure that power is conserved when sensor data is not actively required, allowing longer usage without compromising data capture fidelity.

\subsection{Mounting and Rigidity} 
Lucia ensures precise alignment between its components through a rigid and durable frame. Key sensors are housed in reinforced areas to minimize the risk of misalignment or damage during use. Each unit undergoes thorough factory calibration, with intrinsic and extrinsic parameters precisely set and documented. This calibration supports high-accuracy data capture across different sessions and environments, essential for applications requiring precise sensor fusion and spatial awareness.

% \subsection{Time and Time-alignment} 
% Accurate timestamping is implemented across all sensors in Lucia, aligning data streams with a unified local time source to support multi-modal data fusion. The device also enables synchronization with external devices and systems, making it feasible to integrate data from multiple Lucia units or other compatible devices within a common temporal framework. This synchronization is crucial for collaborative applications and scenarios involving complex data integration.

\subsection{User Interaction and Device Connectivity}

Lucia enables seamless interaction between users and its AI capabilities through multiple methods: hand gestures, voice input akin to Siri, direct input via the hub's interface, and remote control using a mobile phone. These diverse interaction options allow users to intuitively control device functions and access AI features on both the hub and connected mobile devices, enhancing the overall user experience.

The hub is equipped with Bluetooth connectivity, allowing it to connect seamlessly with other wearable smart devices such as smartwatches, AR glasses, and wireless earphones. This feature supports a versatile range of inter-device connections, enabling the hub to control other smart devices and, with appropriate permissions, allowing those devices to control the hub as well.

\section{Software}

\subsection{Operating System}

The hub operates on Android 14, providing a flexible and familiar environment for application development and customization. This platform allows developers to create and install custom programs and extensions, leveraging the extensive Android ecosystem. The operating system supports a wide range of programming languages and frameworks, facilitating the development of sophisticated AI algorithms and user interfaces tailored to specific applications.

\subsection{Data Storage}

Lucia offers robust data storage and connectivity to ensure seamless data management.
Integrated Wi-Fi, Bluetooth, and a USB Type-C port enable real-time data transfer between the device and the user's mobile phone or other devices. This facilitates live streaming of sensor data and remote control functionalities. 
These connectivity features enable real-time data synchronization and offloading, allowing users to store long recordings on their mobile devices or cloud services. This is essential for applications requiring extensive data analysis and archival.

\subsection{AI Algorithms and Models}

Lucia is designed to support advanced AI algorithms and machine learning models, which can be executed either on the hub, the user's mobile phone, or a combination of both, depending on computational requirements and resource availability.

The hub's processing unit is capable of running lightweight AI models locally, enabling real-time data processing and inference without reliance on external devices. 
It has a Neural Processing Unit (NPU) capable of running small, quantized language models. It supports up to 1 TOPS (Tera operations per second) for INT8 operations, 0.5 TOPS for FP16 operations, and 0.25 TOPS for FP32 operations.
For more computationally intensive tasks, data can be offloaded to the user's mobile phone or cloud services, where more powerful processors can handle complex AI workloads.

The device supports popular AI and machine learning frameworks such as TensorFlow Lite and PyTorch Mobile, allowing developers to deploy custom models efficiently.
An SDK and API are provided to facilitate the development and integration of AI applications, including sample code and documentation to accelerate the development process.

By providing flexible options for AI processing, Lucia enables a wide range of applications, from real-time object recognition and activity analysis to personalized assistant functions like question-answering that enhance the user's cognitive capabilities through contextual understanding.

\section{Decoupling Perception and Computation}
Designing an all-day wearable computer with perceptual capabilities presents a challenging trade-off between battery life and form factor. Devices like HoloLens, Quest, and Vision Pro struggle to achieve long-term wearability, while lighter formats like Project Aria offer only data sensing due to power limitations. To enable a computer that supports prolonged wearability while providing adequate computing power, we have decoupled sensing from computing. In our approach, the wearable (the "pin") is dedicated to sensing, while a separate "hub" performs the computational processing.

\section{Privacy Considerations}
\label{sec:privacy_considerations}

User data is handled with the highest level of care, with privacy prioritized through on-device processing and secure data management protocols. 

The camera module is equipped with a ``Privacy Switch'', which allows users to pause video recording effortlessly. An LED light on the module provides a clear visual indicator whenever recording is active, enabling users to monitor and control recording status at any moment.

All data is stored and processed either directly on the device or on the user’s mobile phone, following strict development guidelines. 
The hub has a powerful NPU capable of running small, quantized language models locally. When larger language models are required, users’ personal data, such as daily video recordings, can be vectorized irreversibly before any cloud transmission, preserving privacy while supporting advanced processing capabilities.

\section{Example Applications}

To illustrate the envisioned capabilities and practical benefits of Lucia, we present a set of potential applications that demonstrate how the device could be integrated into various aspects of daily life. 

\subsection{Automatic Activity Recording and Analysis}

Lucia is envisioned to seamlessly integrate into the user's daily life by continuously recording video of their surroundings and activities. Utilizing advanced AI algorithms, the device could automatically recognize and categorize different types of activities throughout the day~\cite{zhang2022deep}. For instance, it may identify when a user is commuting, attending meetings, engaging in physical exercise, socializing with friends, or relaxing at home. The AI could segment the day's events into meaningful categories and timelines, providing a structured overview of the user's daily routines.

This automatic activity recording and analysis has the potential to offer users valuable insights into how they spend their time~\cite{zhou2008activity, pareek2021survey, bukht2024review}. By reviewing summarized activities, users might identify patterns or habits they wish to change, thereby enhancing productivity and personal well-being. For example, if the device indicates that a significant portion of the day is spent on low-priority tasks, the user might adjust their schedule to focus more on important goals. Similarly, recognizing periods of inactivity could encourage the user to incorporate more physical activity into their routine.

Moreover, the device could allow users to set personal goals and track progress over time. By comparing activity levels across days or weeks, users could monitor improvements or identify areas needing attention. The automatic logging removes the need for manual entry, making it a convenient tool for self-improvement and time management.

\subsection{Mood and Behavior Tracking}
Leveraging the device's capability to analyze visual and auditory data, Lucia could potentially offer mood and behavior tracking features. By interpreting cues such as facial expressions, voice tone, speech patterns, and body language, the device may infer the user's emotional state throughout the day~\cite{mehendale2020facial, chamishka2022voice, lu2024gpt}. Advanced algorithms could analyze subtle changes in demeanor, capturing fluctuations in mood that might otherwise go unnoticed~\cite{zhang2020multimodal, lin2021looking, wu2023self, yang2024emollm}.

This information could help users become more aware of their emotional patterns and triggers, facilitating better mental health management. For instance, by identifying correlations between certain activities or environments and mood changes, users might gain insights into factors that contribute to stress or happiness. The device could present this data through intuitive visualizations, showing trends over time and highlighting significant variations.

% If the device detects signs of prolonged stress, anxiety, or sadness, it might offer gentle prompts or suggestions. These could include recommending mindfulness exercises, encouraging breaks, or suggesting relaxation techniques. In some cases, the device might propose reaching out to supportive contacts or provide links to professional mental health resources. By offering timely, personalized support, the device aims to promote emotional well-being.

\subsection{Contextual Question Answering Over Stored Experiences}

One of the most compelling potential applications of Lucia is enabling users to query their recorded experiences through natural language interaction. By leveraging advanced AI models, the device could understand and process user questions, searching through stored data to provide relevant answers~\cite{shen2023encode, he2024malmm, song2023moviechat, song2024moviechat+, ataallah2024goldfishvisionlanguageunderstandingarbitrarily}. Users might ask questions like, "Where did I leave my keys this morning?" or "What was the main topic discussed in yesterday's meeting?" The device could analyze the query, locate pertinent moments in the recorded video, and present concise answers or relevant video snippets.

This contextual question answering could significantly enhance memory recall, serving as an externalized, searchable memory bank~\cite{bermejo2020vimes}. In professional settings, users might retrieve details from meetings, such as agreements made or tasks assigned, improving efficiency and reducing the cognitive load associated with remembering numerous details.

\subsection{Health and Fitness Monitoring}
Lucia could also serve as a comprehensive health and fitness monitoring tool, offering users a convenient way to manage their physical well-being~\cite{tlili2021real,shajari2023emergence, figueira2024wearables}. By automatically detecting physical activities such as walking, running, cycling, or other exercises, the device may help users track their fitness routines without manual input. Using advanced sensors and AI algorithms, it might capture metrics like duration, intensity, distance, and calorie expenditure, providing a detailed overview of physical activity levels.

In addition to tracking activities, the device could analyze movement patterns and behaviors throughout the day. For example, it might monitor posture during sitting or standing, alerting the user to potential ergonomic issues~\cite{stefana2021wearable, tlili2021real,shajari2023emergence, figueira2024wearables}. By recognizing periods of prolonged inactivity, the device could prompt the user to take short breaks, stretch, or engage in light exercise, promoting healthier habits.

\subsection{Safety Applications}

% In terms of safety and security, Lucia could potentially offer features that provide users with added peace of mind in various situations. The continuous recording capability means that, in the event of an accident or unexpected incident, the device might capture valuable footage that could assist in recalling details or providing evidence if needed. For example, if a user is involved in a traffic accident, the recorded video could help reconstruct the event for insurance or legal purposes.

Powered by the AI models behind the device, the device could also enhance personal safety by recognizing potentially hazardous environments or situations. For instance, if the device detects that the user is in an area associated with safety risks, it might send a cautionary alert or suggest alternative routes. Integration with public safety information could enable the device to notify users of nearby incidents or emergencies, helping them stay informed and make safer choices~\cite{karlsson2022visual, shah2024artificial}.

\subsection{Navigation}

With built-in magnetic sensors, Lucia can support precise navigation, particularly indoors where GPS struggles. These sensors detect geomagnetic fields, enabling the device to act as a digital compass and guide users in complex indoor spaces like hospitals, malls, airports, and office buildings.

This functionality could offer turn-by-turn directions, helping users quickly find specific rooms, stores, or exits without needing to rely on handheld devices~\cite{anderson2018vision, gu-etal-2022-vision, zhang2024vision}. In professional settings, it can streamline navigation for employees or visitors in large facilities. Additionally, Lucia’s navigation support enhances augmented reality experiences by aligning virtual directions with real-world paths, making orientation and wayfinding easier and more intuitive.

\section{Conclusion}
Lucia introduces a breakthrough in personalized AI through Temporal Computing, creating a device that understands and adapts to users over time. By combining all-day comfort with high perceptual capabilities, the device offers continuous, real-time data access in an unobtrusive design, setting it apart from current devices that often lack this balance.
This report has highlighted Lucia’s potential to transform human-computer interaction by enhancing memory recall, decision-making, and productivity in a seamless way. By sharing this open-source platform with research institutions, we hope to inspire advancements in personalized AI and contribute to a more interconnected, intelligent future.

\section*{Ethical Statement}
The device developed in this project is designed with the highest possible level of data protection, as introduced in Sec.~\ref{sec:privacy_considerations}. However, as the platform is open to developers for validating and testing contextual intelligence applications, users should remain mindful of potential risks when installing applications and granting data permissions to third-party developers.

% \section*{Acknowledgments}

\bibliography{custom}

\appendix

\section{Detailed Specifications}
\label{sec:appendix:specifications}

The detailed specifications of the device are provided in Table~\ref{tab:specifications}.

\begin{table*}[!htp]
\centering
\footnotesize
\resizebox{\textwidth}{!}{

\begin{tabular}{|p{2cm}|p{4cm}|p{8cm}|}
\hline
\textbf{Category} & \textbf{Specification} & \textbf{Lucia} \\
\hline
\multirow{25}{*}{\shortstack{Camera \\ (RV1106g3)}} & Sensor (CMOS) & IMX681 \\
\cline{2-3}
 & Sensor Resolution & 12M \\
\cline{2-3}
 & Video Format & 2016x1512@24/25/30fps \newline
1920x1080@24/25/30fps \newline
1600x1200@24/25/30fps \newline
1280x720@24/25/30fps \newline
1024x768@24/25/30fps \newline
640x480@24/25/30fps \newline
H.264
\\
\cline{2-3}
 & Photo Format & 3840x2160~(16:9)\newline
2016x1512~(4:3)\newline
1920x1080~(16:9)\newline
1600x1200~(4:3)\newline
JPG \\
\cline{2-3}
 & FOV & 94°x77° \\
\cline{2-3}
 & Aperture & F2.25 \\
\cline{2-3}
 & Main Chip & Rockchip RV1106G2 \\
\cline{2-3}
 & Memory Size & 128MB DDR3 \\
\cline{2-3}
 & EMMC Size & 32GB \\
\cline{2-3}
 & MIC & Dual MIC \\
\cline{2-3}
 & Speaker & Single speaker 1W \\
\cline{2-3}
 & Buzzer & No \\
\cline{2-3}
 & Gyroscope & 6-axis \\
\cline{2-3}
 & Battery Capacity & 800mAh \\
\cline{2-3}
 & Bluetooth & BLE5.0 \\
\cline{2-3}
 & WIFI & 2.4GHz; 802.11a/b/g/n/ac \\
\cline{2-3}
 & Bluetooth Earphone & Supported \\
\cline{2-3}
 & Working Time & ~160 min \\
\cline{2-3}
 & Standby Time & / \\
\cline{2-3}
 & Output Interface & WiFi Wireless/Bluetooth/USB \\
\cline{2-3}
 & Indicator Light & Breathing light effect \\
\cline{2-3}
 & Privacy Switch & Yes \\
\cline{2-3}
 & Ambient Light Sensor & Yes \\
\cline{2-3}
 & IR Supplementary Light & Yes \\
\cline{2-3}
 & Front and Rear Camera Insertion & Pending \\
\cline{2-3}
 & Camera Dimensions & 60×38.5×9mm \\
\cline{2-3}
 & Camera Weight & 44 grams \\
\cline{2-3}
 & Shell Material & Metal frame, dual glass \\
\cline{2-3}
 & Working Temperature & -10°C to 50°C \\
\hline
\multirow{16}{*}{HUB (Rk3562)} & Screen Size & 4.1" \\
\cline{2-3}
 & Screen Resolution & 720X1280 \\
\cline{2-3}
 & Main Chip & RK3562 \\
\cline{2-3}
 & Memory Size & 8GB \\
\cline{2-3}
 & EMMC Size & 128GB \\
\cline{2-3}
 & MIC & Dual MIC \\
\cline{2-3}
 & Speaker & Single speaker 1W \\
\cline{2-3}
 & Bluetooth & BLE5.0 \\
\cline{2-3}
 & WIFI & 2.4GHz; 802.11a/b/g/n/ac \\
\cline{2-3}
 & Bluetooth Earphone & Supported \\
\cline{2-3}
 & Battery Capacity & 4000mAh \\
\cline{2-3}
 & Working Time & ~400 min \\
\cline{2-3}
 & Standby Time & ~10 days \\
\cline{2-3}
 & Output Interface & WiFi Wireless/Bluetooth/USB (Type-C) \\
\cline{2-3}
 & Wireless Charging & Supported \\
\cline{2-3}
 & HUB Dimensions & 60×114×12mm \\
\cline{2-3}
 & HUB Weight & / \\
\cline{2-3}
 & Operating System & Android 14 \\
\cline{2-3}
 & Shell Material & Metal frame, glass back \\
\cline{2-3}
 & Working Temperature & -10°C to 50°C \\
\hline
\end{tabular}

}
\caption{Specifications of the device.}
\label{tab:specifications}
\end{table*}

\end{document}